\documentclass[%
 reprint,
superscriptaddress,
 amsmath,amssymb,
 aps,
 prl
]{revtex4-2}

\usepackage{graphicx}
\usepackage{dcolumn}
\usepackage{bm}
\usepackage[dvipsnames]{xcolor} 
\usepackage[english]{babel}

\begin{document}
\selectlanguage{english}

\preprint{APS/123-QED}

\title{
Stress Relaxation Driven by Elementary Plastic Events in Non-Ageing Liquid Foams
}

\author{Florian Schott}
\email{florian.schott@solid.lth.se}
\affiliation{Division of Solid Mechanics, LTH, Lund University, Lund, Sweden.}
\author{Benjamin Dollet}%
\affiliation{ Université Grenoble Alpes, CNRS, LIPhy, 38000 Grenoble, France}
\author{Christian M. Schlepütz}%
\affiliation{Swiss Light Source, Paul Scherrer Institut, 5232 Villigen, Switzerland}
\author{Cyrille Claudet}%
\affiliation{Université Côte d'Azur, CNRS, INPHYNI, France}
\author{Stefan Gstöhl}%
\affiliation{Swiss Light Source, Paul Scherrer Institut, 5232 Villigen, Switzerland}
\author{Rajmund Mokso}%
\affiliation{Department of Physics, Technical University of Denmark}
\affiliation{Division of Solid Mechanics, LTH, Lund University, Lund, Sweden.}
\author{Stéphane Santucci}%
\affiliation{ Laboratoire de Physique, UMR CNRS 5672, ENS de Lyon, Université de Lyon, Lyon, France}
\author{Christophe Raufaste}%
\email{christophe.raufaste@univ-cotedazur.fr}
\affiliation{Université Côte d'Azur, CNRS, INPHYNI, France}
\affiliation{Institut Universitaire de France (IUF), France}

\date{\today}

\begin{abstract}
Liquid foams are archetypal athermal  amorphous solids whose elasticity arises from the jamming of densely packed bubbles. We investigate the stress relaxation of non-ageing liquid foams following flow cessation, using fast X-ray tomo-rheoscopy. Thanks to \emph{in situ}, time-resolved measurements, we uncover robust linear affine relationships between shear stress, plastic activity, and coordination number throughout the relaxation toward a residual stress state below the yield value. In contrast to previous  studies on amorphous solids, we observe an exponential relaxation governed by the duration of individual plastic events, rather than by cascades of correlated ones associated with much longer, shear-rate–dependent timescales or power-law relaxations. 
Our results provide new insight into the microscopic mechanisms governing the time evolution of stress in jammed amorphous solids.
\end{abstract}

\maketitle

Complex materials such as foams, creams, pastes, and concretes behave as solids under small deformations, with elasticity arising from a disordered microstructure of densely packed elements \cite{balmforth_yielding_2014, coussot_yield_2014, bonn_yield_2017, nicolas_deformation_2018, nelson_designing_2019}. These jammed amorphous solids flow above a yield stress, driven by shear-induced structural rearrangements. Following flow cessation, they retain a residual stress below the yield stress, which depends on the pre-shear rate $\dot{\gamma}$, with higher rates leading to lower residual stress \cite{Ballauff2013, Mohan2013, Mohan2015, Moghimi2017, zakhari_stress_2018, Vasisht2022, vinutha_memory_2024}.
This intriguing behavior reflects the material’s memory of its shear history \cite{vinutha_memory_2024} and highlights the challenges involved in establishing shear protocols for preparing samples with isotropic mechanical properties \cite{bertrand_protocol_2016, choi_optimal_2020, edera_mechanical_2025, beyer_shear-induced_2025, blanc_disorder-induced_2026}.

The existence of a residual stress following flow cessation requires the system to remain stable against ageing, i.e., against long-time structural evolution~\cite{nicolas_deformation_2018}, as observed in athermal particle-based simulations~\cite{Vasisht2022, vinutha_memory_2024}, experiments on suspensions of micron-sized particles~\cite{vinutha_memory_2024}, and athermal mesoscale elastoplastic simulations~\cite{Vasisht2022}. Colloidal systems may still undergo ageing, but residual stresses can nevertheless be identified when the stress relaxation occurs on timescales much shorter than those associated with ageing~\cite{Ballauff2013, Mohan2013, Mohan2015, Moghimi2017, zakhari_stress_2018}. 
Whenever the microscopic relaxation time of individual structural rearrangements was reported \cite{Mohan2013, Mohan2015, Vasisht2022, vinutha_memory_2024}, the overall relaxation dynamics was found to occur on much longer time scales (although still shorter than those associated with ageing if any), evidencing strong temporal correlations. This feature is embedded in most elastoplastic models, in which local plastic events induce stress redistribution in their surroundings, triggering others nearby and successive avalanches \cite{nicolas_deformation_2018}. While each individual event relaxes on the microscopic time, the cascade of correlated events relaxes on a much longer scale. Moreover, the smaller the pre-shear rate $\dot{\gamma}$, the slower the resulting dynamics \cite{Mohan2013, Mohan2015, Vasisht2022, vinutha_memory_2024}, consistent with the increase in correlation length observed at decreasing shear rates~\cite{vinutha_memory_2024}.

Among jammed amorphous solids, liquid foams consist of dispersions of gas bubbles in a liquid phase~\cite{cantat2013foams}. Usual foams coarsen~\cite{saint-jalmes_physical_2006}, whereby gas exchange between bubbles progressively alters the foam microstructure and induces rearrangements. This ageing process leads to a gradual loss of rigidity~\cite{hohler_rheological_1999, gopal_relaxing_2003}, fluid-like behavior in creep and recovery tests~\cite{cohen-addad_origin_2004, marze_protein_2005, cantat_mechanical_2005}, power-law rheology~\cite{cates_tensorial_2004, hwang_understanding_2016, lavergne_delayed_2022, rodriguez-cruz_experimental_2023, thirumalaiswamy_slow_2025} and plays a role analogous to an effective temperature in glassy materials \cite{sollich_rheology_1997, fielding_aging_2000}. 
With the recent development of ultrastable foams~\cite{rio_unusually_2014, Bergeron22}, and with typical bubble sizes well above one micron, coarsening and other ageing phenomena can be strongly delayed, making liquid foams an experimental model system for studying flow cessation in athermal, non-ageing amorphous solids. 
Importantly, they are also unique in that their internal structure and mechanical state can be directly observed in space and time using tomo-rheoscopy \cite{Schott2025}, a non-intrusive technique recently employed to measure the relaxation time of individual plastic events and the associated stress redistribution in their surroundings. 

In this Letter, we investigate the relaxation dynamics following flow cessation in non-ageing liquid foams subjected to different pre-shear rates. 
First, enabled by simultaneous intrinsic measurements of shear stress, plastic activity, and coordination number, we demonstrate linear affine relationships between these local quantities during relaxation, with coefficients that depend on the pre-shear history. 
Second, in contrast to results obtained in other systems \cite{Mohan2013, Mohan2015, Vasisht2022, vinutha_memory_2024}, the relaxation time remains of the order of the microscopic relaxation time. This indicates that the dynamics is not governed by correlations or nonlocal effects \cite{goyon_spatial_2008, bocquet_kinetic_2009, mansard_local_2012, jop_microscale_2012, ferrero_relaxation_2014, lin_scaling_2014, benzi_unified_2019, clemmer_criticality_2021, vinutha_memory_2024}. Our results thus show that stress relaxation below the yield stress can arise from plasticity without triggering long-lasting cascades of plastic events.
These results open a new route for improving elastoplastic models of soft jammed solids. 

\begin{figure}[b!]
    \centering
    \includegraphics[width=0.7\linewidth]{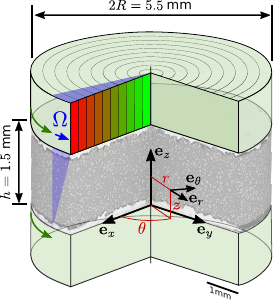}
    \caption{
    Experimental setup.  
    In the plate–plate geometry, the shear rate in the pre-shear phase varies with radius according to $\dot{\gamma}(r) = \Omega r / h$, where $\Omega$ is the relative rotation speed. 
    Measurements are performed within toroidal regions at various radii and corresponding shear rates, with color coding from green ($r = 0.1375$~mm, $\dot{\gamma} = 0.0006$~s$^{-1}$, near the center) to red (2.6125 mm, 0.0109 s$^{-1}$, at the outer edge). 
    Typical movies of the ($\theta$ - $z$) plane at mid-radius ($r = 1.375$ mm) are provided in the Supplemental Material \cite{SuppMat}. 
    }
    \label{fig:fig1}
\end{figure}


Liquid foams that are stable against coalescence, coarsening, and drainage~\cite{rio_unusually_2014} were prepared as follows.  
A surfactant solution was made according to the protocol of Golemanov et al. \cite{golemanov_surfactant_2008}. First, 6.6 wt\% sodium lauryl ether sulfate (SLES) and 3.4 wt\% cocamidopropyl betaine (CAPB) were mixed in ultrapure water. Then, 0.4 wt\% myristic acid (MAc) was added and dissolved by stirring and heating the mixture at 60$^\circ$C for one hour. This stock solution was subsequently diluted 20-fold with a 50/50 (by mass) glycerol/water mixture to obtain the final solution. Bubbles were generated using a co-flow microfluidic device, by simultaneously injecting the solution and perfluorohexane-saturated air. Control of the liquid flow rate and gas pressure was achieved using a Harvard Apparatus PHD Ultra syringe pump and a Fluigent MFCS-FLEX pressure controller, respectively. This setup produced bubbles with mean radii between 54 and 87 $\mu$m and low size dispersity. Bubbles aggregate at the outlet of the device to form a wet foam. To reduce the liquid fraction, the foams were subsequently centrifuged at various speeds using an Eppendorf 5702 centrifuge. The final liquid fraction $\phi_\ell$ ranged from 8.4 \% (driest) to 21.7 \% (wettest). These foams were characterized in Schott et al. \cite{Schott2025}, which confirmed that their properties remained unchanged throughout each experiment.

Samples were probed using tomo-rheoscopy at the TOMCAT beamline of the Paul Scherrer Institute (Switzerland). A dual-motor rheometer (Anton Paar MCR 702 TwinDrive) was configured in a plate–plate geometry with a 5.5 mm diameter and a 1.5 mm gap, as detailed in~\cite{Schott2025}. To ensure comparable deformation histories across experiments, foam samples were first sheared at a relative rotation speed of 50 mHz for 20 s in one direction, followed by 20 s in the opposite direction. After a rest period of 80 s, the foam was sheared at 1 mHz for 480 s. Finally, shear was stopped and relaxation was recorded over 60 s. The plate–plate geometry allows the same foam sample to be probed under different shear rates simultaneously, as the pre-shear rate $\dot{\gamma}$ just before flow cessation varies linearly with radial position. As verified in \cite{Schott2025}, there is no slip at the plates and the shear rate is homogeneous at a given radius, reaching a maximum of 0.0116 s$^{-1}$ at the periphery. Three-dimensional images were acquired every 3 s, covering a volume of $2.2 \times 5.5 \times 5.5$ mm$^3$ with a voxel size of 2.75 $\mu$m, and were analyzed as described in \cite{Schott2025}. The raw grayscale images depict darker regions corresponding to the gas phase and lighter regions to the liquid phase. Because the thin films between adjacent bubbles are below the resolution limit, the foams appear as open-cell structures. Image processing enabled the reconstruction and labeling of individual bubbles. For each bubble, we extracted its shape, position, contact topology with neighbors, and stress \cite{schott_foamquant_2025}. 
As described in our previous study \cite{Schott2025}, in the regime where viscous stresses are negligible, local elastic stress can be computed from the geometry of individual bubble interfaces and knowing the value of the surface tension.

We first focus on two quantities: the shear stress  and the plastic activity. The stress is computed by averaging the stress tensors of individual bubbles whose centroids lie within defined sub-volumes. Due to the axisymmetric geometry, stress is expressed in cylindrical coordinates within concentric tori (see Fig.~\ref{fig:fig1}). Each torus encompasses all bubbles located between radial positions $r$ and $r + \Delta r$. In practice, the radial domain is divided into 10 segments, each typically containing between 200 and 5000 bubbles, from the innermost to the outermost regions. Given the geometry, the relevant component of the stress tensor is the $\theta$–$z$ (azimuthal–vertical) component, denoted $\sigma$. Plastic activity $P$ within the same tori is defined as the number of detected topological events—identified as the formation or loss of a contact between two bubbles—between two time steps, normalized by the number of bubbles in the corresponding torus. 

We denote by $\sigma_{ss}$ and $P_{ss}$ the average steady-state values of the shear stress and plastic activity, respectively, measured during the pre-shear phase preceding flow cessation. As shown in Fig.~\ref{fig:app:sigma_gamma} in End Matter, the data obtained at the lowest shear rate did not reach the plastic plateau and were therefore excluded from the analysis. 
Data from the outermost torus at $\dot{\gamma} = 0.0109$ s$^{-1}$ were also excluded due to image-analysis artifacts near the field edges. 
As described in \cite{Schott2025}, for each experiment, $\sigma_{ss}$ increases with the shear rate $\dot{\gamma}$, and the yield stress is inferred from the extrapolation to $\dot{\gamma} = 0$ (Fig.~\ref{fig:app:steady_state_sig} in End Matter). The plastic activity $P_{ss}$ also increases with $\dot{\gamma}$ within each experiment, following an approximately proportional trend (Fig.~\ref{fig:app:steady_state_P} in End Matter). When comparing different experiments, the proportionality coefficient between $P_{ss}$ and $\dot{\gamma}$ increases with the liquid fraction, indicating that wetter foams undergo more structural rearrangements than drier foams under the same imposed strain.

\begin{figure}[t!]
    \centering
    \includegraphics[width=1\linewidth]{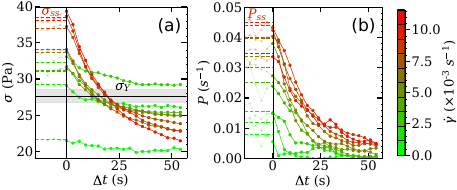}
    \caption{
(a) Shear stress $\sigma$ and (b) plastic activity $P$ relaxation curves as functions of the relaxation time $\Delta t$ for the reference experiment. The color scale, from green to red, represents the local shear rate prior to flow cessation. $\sigma_{ss}$ and $P_{ss}$ are indicated for the largest $\dot{\gamma}$. The yield stress $\sigma_Y$ is shown as a horizontal line, with the gray band representing its standard deviation, indicating that most stress curves relax below $\sigma_Y$.
    }
    \label{fig:fig2}
\end{figure}

\emph{}

Figure~\ref{fig:fig2} shows typical measurements for a representative experiment, with color coding for the pre-shear rate (or equivalently, the radial position from the axis of rotation), covering the last 15 s of the pre-shear phase and the subsequent relaxation starting at $\Delta t = 0$. Regardless of the pre-shear rate $\dot{\gamma}$, the shear stress $\sigma(\Delta t)$ decreases from its steady-state value $\sigma_{ss}(\dot{\gamma})$. It is not immediately evident that $\sigma$ saturates to a residual value  (we will confirm this below). During this relaxation $\sigma$ drops below the yield stress $\sigma_Y$, except for one torus, where limited statistical sampling at small inner radii leads to significant fluctuations. Moreover, the higher the pre-shear rate, the larger the  decrease in stress, leading to a crossing of the stress relaxation curves for different $\dot{\gamma}$ values. The saturation behavior is more clearly seen in the evolution of the plastic activity $P(\Delta t)$, which begins at $P_{ss}(\dot{\gamma})$ and shows a marked decay toward zero. 

\begin{figure}[b!]
    \centering
    \includegraphics[width=1\linewidth]{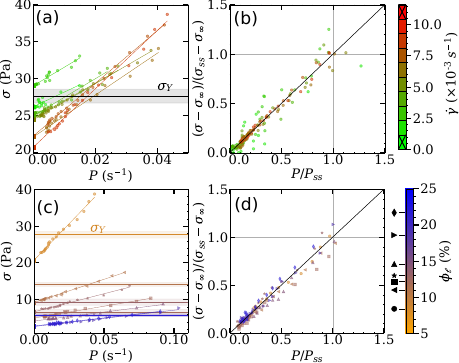}
    \caption{
(a) Shear stress $\sigma$ as a function of plastic activity $P$ in the reference experiment (data of Fig.~\ref{fig:fig2} excluding the innermost and outermost tori), showing affine relations for each shear rate. Solid lines are affine fits, whose intercepts define the $\sigma_{\infty}$ values, as used in Eq.~\eqref{eq:stressplasticity}. 
(b) Normalized $\sigma$ versus normalized $P$ for the same dataset.  
(c) $\sigma$ as a function of $P$ for different liquid fractions at a fixed shear rate, $\dot{\gamma} = 0.0098$ s$^{-1}$. Symbols associated with each liquid fraction are shown next to the lower color bar. 
(d) Normalized $\sigma$ versus normalized $P$ for the same dataset. 
    }
    \label{fig:fig3}
\end{figure}
Strikingly, shear stress and plastic activity are closely correlated during relaxation: plotting $\sigma$ as a function of $P$—both treated as parametric functions of time— shows affine relationships, with both the slope and intercept varying with the pre-shear rate (Fig.~\ref{fig:fig3}(a)). For each $\dot{\gamma}$ within a given experiment, the residual stress $\sigma_\infty(\dot{\gamma})$ corresponds to the intercept—that is, the stress value when plastic activity vanishes. This strong correlation yields the data collapse shown in Fig.~\ref{fig:fig3}(b), and
the relation :
\begin{equation}\label{eq:stressplasticity}
\frac{\sigma(\Delta t) - \sigma_\infty(\dot{\gamma})}{\sigma_{ss}(\dot{\gamma}) - \sigma_\infty(\dot{\gamma})}  = \frac{P(\Delta t)}{P_{ss}(\dot{\gamma})} .
\end{equation}
Such relationship was further tested using foams with various liquid fractions. For simplicity, Fig.~\ref{fig:fig3}(c) shows data at a single radial position corresponding to a shear rate of $\dot{\gamma} = 0.0098$ s$^{-1}$. Again, we observe a clear affine relationship between shear stress $\sigma$ and plastic activity $P$, allowing for an unambiguous determination of the residual stress $\sigma_\infty$ for each experiment at this shear rate. This measurement, together with the data collapse of the normalized stress versus the normalized plastic activity, shown in Fig.~\ref{fig:fig3}(d), confirms the validity of Eq.~\eqref{eq:stressplasticity} for experiments performed on various liquid foams. 
The constancy of the slope $d\sigma/dP$ indicates that, on average, plastic events produce a stress variation proportional to their amplitude, for a given pre-shear rate and liquid fraction. We found that this is indeed the case during the pre-shear phase, where the statistics are sufficient to quantify this relationship~\cite{Schott2025}, and it appears to remain valid during the relaxation phase.

\begin{figure}[t!]
    \centering
    \includegraphics[width=1.0\linewidth]{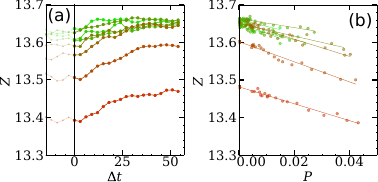} 
\caption{(a) Mean coordination number $Z$ as a function of relaxation time $\Delta t$ for various shear rates in the reference experiment. 
(b) $Z$ as a function of the plastic activity $P$ for the same experiment, revealing affine relations for each shear rate. Solid lines are best fits, with intercepts defining the residual coordination number $Z_\infty$. 
Same color coding~as~in~Fig.~\ref{fig:fig3}.
}
\label{fig:fig4new}
\end{figure}

Further insights can be obtained by analyzing structural properties such as the mean coordination number $Z$, which has attracted a lot of attention in the context of the jamming of soft materials, including foams \cite{Katgert2013}. 
Notably, concomitant with the reduction in shear stress and plastic activity, the foam  exhibits a systematic increase in $Z$, reflecting a progressive consolidation of its microstructure throughout the relaxation process (Fig.~\ref{fig:fig4new}(a)). This behavior agrees with observations reported in numerical simulations of other soft jammed solids~\cite{Mohan2013,vinutha_memory_2024}, but to our knowledge, it has not been reported in experiments. Indeed, the existing experimental measurements of the coordination number, notably its dependence on the liquid fraction, are in the static regime \cite{Jorjadze2013}. 
Interestingly, larger pre-shear rates $\dot{\gamma}$ correspond to lower $Z$ values in the pre-shear phase, consistent with a loss of elastic integrity in the contact network~\cite{ngouamba_elastoplastic_2019, cuny_microscopic_2021, cuny_dynamics_2022, vinutha_stressstress_2023}. 
More importantly, once again, an affine relation with the microscopic plastic activity is found (Fig.~\ref{fig:fig4new}(b)), allowing us to define a residual coordination number $Z_{\infty}$, and to further highlight a relation between the normalized coordination number $(Z-Z_{\infty})/(Z_{ss}-Z_{\infty})$ and the normalized plastic activity $P/P_{ss}$ (data collapse shown in Fig.~\ref{fig:app:normZ_normP}(a) of the End Matter).  
Plotting $Z_{\infty}$ as a function of $\dot{\gamma}$ for foams with different liquid fractions reveals that the residual coordination number remains essentially constant, depending primarily on the liquid fraction (Fig.~\ref{fig:app:normZ_normP}(b) in End Matter).  
For the driest foam, $Z_{\infty}$ is close to 13.7, the value expected for disordered monodisperse foams in the dry limit~\cite{kraynik_structure_2003}, and drops below 12 for the wettest foam. 
This indicates that the microstructure relaxes toward a network characteristic of the foam’s liquid content. 
However, in the reference experiment with a dry foam, the network does not fully recover at the largest shear rates, showing that memory effects can also be encoded in structural quantities under certain conditions. 
Nevertheless, this contribution remains secondary to the predominant influence of the liquid fraction.

\begin{figure}[b!]
    \centering
    \includegraphics[width=1\linewidth]{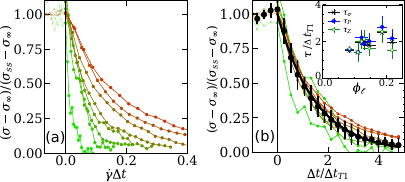}
    \caption{
Normalized shear stress, $(\sigma - \sigma_\infty)/(\sigma_{ss} - \sigma_\infty)$, as a function of $\dot{\gamma} \Delta t$  (a) and $\Delta t/\Delta t_{T1}$ (b) for the reference experiment. Same color scale as in Fig.~\ref{fig:fig3}. 
In (b), the black dots correspond to averages over all shear rates, fitted by an exponential decay of the form $e^{-\Delta t/\tau}$ with $\tau = (1.50 \pm 0.13)\,\Delta t_{T1}$ (thick black line).
Inset of (b): $\tau/\Delta t_{T1}$ as a function of $\phi_\ell$, with the three data points for each experiment obtained from fits of the normalized stress, plastic activity, and mean coordination number.
    }
    \label{fig:fig5new}
\end{figure}

Finally, we examine the temporal evolution of the stress during relaxation. By construction, the normalized stress defined in Eq.~\eqref{eq:stressplasticity} starts at unity and decays to zero as the system relaxes. 
Plotting the normalized stress as a function of $\dot{\gamma} \Delta t$ for the reference experiment results in a spreading of the data (Fig.~\ref{fig:fig5new}(a)), showing that $1/\dot{\gamma}$ is not the relevant timescale of relaxation in our experiments. 
In contrast, Fig.~\ref{fig:fig5new}(b) shows the normalized stress as a function of $\Delta t/\Delta t_{T1}$, where $\Delta t_{T1}$ is the typical duration of a single structural rearrangement, called T1, as characterized in Ref.~\cite{Schott2025}. This timescale reflects the microscopic relaxation time of a plastic event and varies weakly between experiments, ranging from 8~s to 12~s from the wettest to the driest foams~\cite{Schott2025}. All data collapse onto a single master curve, with a stress relaxation of 50\% reached at $\Delta t \simeq \Delta t_{T1}$, showing that $\Delta t_{T1}$ is the relevant timescale governing the dynamics. 
In addition, the master curve exhibits a behavior compatible with an exponential relaxation over at least one order of magnitude. This trend is particularly evident in log-linear representations (Fig.~\ref{fig:app:semilog}(a) in the End Matter) and can be fitted by $e^{-\Delta t/\tau}$, where $\tau$ is the characteristic relaxation time. Consistent with the intrinsic affine relationships, the plastic activity and the mean coordination number follow the same trend (Figs.~\ref{fig:app:PPss_DtDtT1}(b) and (c) in the End Matter). 
Analyzing the other experiments performed at different liquid fractions, we observe the same behavior, with $\tau$ remaining of the order of the microscopic timescale and the ratio $\tau/\Delta t_{T1}$ showing no significant dependence on the liquid fraction (inset of Fig.~\ref{fig:fig5new}(b)).

This indicates that stress relaxation after flow cessation is primarily governed by the plastic events triggered immediately upon cessation, in contrast with previous studies  \cite{Mohan2013, Mohan2015, Vasisht2022, vinutha_memory_2024}. In those systems, cascades of correlated plastic events give rise to relaxation timescales much longer than the microscopic one. Dimensional analysis suggests that the dynamics could be described by the variable $\dot{\gamma}^{\alpha} \Delta t/\Delta t_{T1}^{1-\alpha}$, where $\alpha\in[0,1]$ quantifies the relative contribution of correlated events: $\alpha=0$ corresponds to dynamics governed by the microscopic timescale, while $\alpha=1$ indicates full dominance of time-correlated plastic events. 
Our system lies close to the former limit, whereas a reanalysis of the experiments by Vinutha et al. \cite{vinutha_memory_2024} reveals the latter (Supplemental Material). 
Surprisingly, particle-based \cite{Mohan2013, Mohan2015, Vasisht2022, vinutha_memory_2024} and mesoscale \cite{Vasisht2022} simulations do not reproduce the experimental observations and lie between the two limits (Supplemental Material), highlighting a broad spectrum of behaviors depending on the systems.

In contrast, our observations could be interpreted within the theoretical framework proposed by Cuny et al. \cite{cuny_microscopic_2021}, which neglects correlated dynamics and identifies the microscopic timescale—here $\Delta t_{T1}$—as the relevant relaxation time, consistent with $\alpha=0$. In this framework, $\dot{\gamma}$–dependent residual stresses below the yield stress arise from tensorial effects: pre-shear modifies the relative proportions of shear and normal stress components, while relaxation reduces the stress amplitude without altering these proportions. Linearization of the relaxation dynamics at $\dot{\gamma}=0$ (Eq.~(14) in \cite{cuny_microscopic_2021}) then predicts an exponential decay, consistent with our observations.


To conclude, first of all, we have uncovered a robust linear affine relationship between microscopic quantities, such as shear stress, bubble coordination number and plastic activity, during flow cessation experiments in non-ageing liquid foams. This raises the question of whether such a relationship might also hold for other soft jammed solids. Although direct experimental verification is currently limited—owing to the challenge of simultaneously measuring stress and plastic activity—reanalyzing numerical data from Vasisht et al. \cite{Vasisht2022}, who simulated flow cessation using athermal mesoscale elasto-plastic models (see Supplemental Material), suggest that this relationship could indeed be a more general feature of jammed amorphous solids. As such, if confirmed, our results would provide a valuable tool for characterizing relaxation dynamics, particularly in simulations where both quantities are readily accessible. 

While our experimental system also exhibits residual stresses below the yield stress, it differs from others in that the relaxation time is set by the typical duration of individual plastic events and consequently shows no dependence on the pre-shear rate. This points to a relaxation dynamics that is not driven by cascades of correlated plastic events. Instead, the presence of a single relaxation time and an exponential stress decay is consistent with a dynamics governed by orientational effects of the stress tensor, as suggested by Cuny et al. \cite{cuny_microscopic_2021}.  
Although this model does not explicitly describe plastic activity, and assumes a simple relaxation dynamics of the stress components above a threshold amplitude of the stress anisotropy, it explains the existence of residual stresses below the yield stress as resulting from a change in the relative contributions of shear and normal components of the stress anisotropy during relaxation. These quantities should be accessible in numerical studies and, importantly, also in our experimental system, which will be the subject of future work. 
In practice, correlated plasticity and orientational effects may coexist and jointly contribute to the emergence of residual stresses below the yield stress, with their relative importance depending on the system considered. 
Future studies should aim to identify the key parameters (deformability of the building elements, nature of the dissipation, temperature in colloidal suspensions ...) that determine whether a system falls within the $\alpha = 0$ or $\alpha = 1$ limits or exhibits intermediate behavior. Reanalysis of the experimental data in \cite{vinutha_memory_2024} (the $\alpha = 1$ limit) reveals that the normalized stress follows a power-law scaling in $\dot{\gamma} \Delta t$ (Supplemental Material), suggesting that exponential versus power-law relaxation may distinguish uncorrelated from correlated dynamics. 
A natural perspective of this work is to use our stress relaxation data as a benchmark for tensorial elastoviscoplastic models that do not rely on correlated plasticity \cite{saramito_new_2007, saramito_new_2009, belblidia_computations_2011, cuny_microscopic_2021, cuny_dynamics_2022}, as well as for their adaptations to liquid foams \cite{marmottant_discrete_2008, raufaste_discrete_2010, cheddadi_understanding_2011}. This will allow us to assess their ability to capture the details of relaxation dynamics and, more generally, the fine rheological features of foams and other yield-stress fluids.

\section*{Acknowledgements}
We wish to thank the Swedish Research Council for funding this project (grant No. 2019-03742). We acknowledge the Paul Scherrer Institut, Villigen, Switzerland for provision of synchrotron radiation beamtime at the TOMCAT beamline X02DA of the SLS. The Tomo-Rheoscope used in this study was funded by the Swiss National Science Foundation (Grant No. 205311, https://data.snf.ch/grants/grant/205311).
S.S. thanks support from the CNRS and ENS de Lyon, through the IRP (D-FFRACT). 
We thank Thibaut Divoux, Sébastien Manneville, Emanuela Del Gado, Kirsten Martens, Romain Mari, Simon Cox and Pascal Hébraud for fruitful discussions. 
We thank H. A. Vinutha, Véronique Trappe and Emanuela Del Gado for sharing their data.

\section*{Data availability}
The data that support the findings of this article are not publicly available. The data are available from the authors upon reasonable request.

\bibliography{relax_bib}


\newpage
\hspace{1cm}
\appendix
\newpage

\section*{END MATTER}

\section{Shear stress versus imposed strain}

\begin{figure}[!h]
    \centering
    \includegraphics[width=1\linewidth]{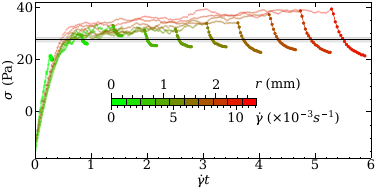}
    \caption{Shear stress $\sigma$ as a function of imposed strain $\dot{\gamma} t$ over the full duration of the reference experiment. The horizontal line indicates the measured yield stress. 
    }
    \label{fig:app:sigma_gamma}
\end{figure}

Figure \ref{fig:app:sigma_gamma} shows the shear stress at all radial positions as a function of the imposed strain $\dot{\gamma} t$ over the full duration of the reference experiment. The innermost region of the foam does not reach the plastic plateau before flow cessation. Therefore, the corresponding shear-rate data were excluded from the analysis in all series.


\section{Steady-state values}

\begin{figure}[!h]
    \centering
    \includegraphics[width=0.7\linewidth]{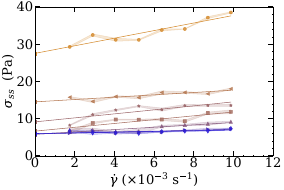} 
    \caption{
Average steady-state shear stress $\sigma_{ss}$ as a function of shear rate $\dot{\gamma}$ for all experiments, measured over the 15 s preceding flow cessation. The yield stress $\sigma_Y$ is obtained by linear extrapolation in the limit $\dot{\gamma} \rightarrow 0$.
}
    \label{fig:app:steady_state_sig}
\end{figure}

\begin{figure}[!h]
    \centering
    \includegraphics[width=1\linewidth]{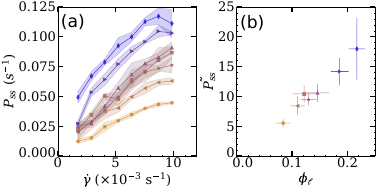}
    \caption{
(a) Average steady-state plastic activity $P_{ss}$ as a function of shear rate $\dot{\gamma}$ for all experiments, measured over the 15 s preceding flow cessation. 
(b) Proportionality coefficient between $P_{ss}$ and $\dot{\gamma}$, $\tilde{P}_{ss}$, as a function of the liquid fraction. 
    }
    \label{fig:app:steady_state_P}
\end{figure}

In the pre-shear phase preceding flow cessation, we measured the average steady-state values of the shear stress and plastic activity, denoted by $\sigma_{ss}$ and $P_{ss}$, respectively. Averages were performed over the 15 s prior to flow cessation. Figure~\ref{fig:app:steady_state_sig} shows $\sigma_{ss}$ as a function of $\dot{\gamma}$. The steady-state stress $\sigma_{ss}(\dot{\gamma})$ exhibits the characteristic behavior of a yield-stress fluid, consistent with the results already reported in Schott et al. \cite{Schott2025}. The yield stress $\sigma_Y$ is obtained by linear extrapolation as $\dot{\gamma} \to 0$.
In Fig.~\ref{fig:app:steady_state_P}(a), $P_{ss}$ is plotted as a function of $\dot{\gamma}$. The steady-state plastic activity increases with $\dot{\gamma}$, following an approximately proportional trend within each experiment. The proportionality coefficient increases with liquid fraction, as shown in Fig.~\ref{fig:app:steady_state_P}(b).

\section{Residual shear stress}

\begin{figure}[h!]
    \centering
    \includegraphics[width=0.7\linewidth]{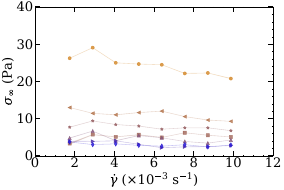} 
    \caption{Residual shear stress $\sigma_{\infty}$ as a function of shear rate $\dot{\gamma}$ for all the experiments obtained from the linear extrapolation method shown in Fig.\ref{fig:fig3}}
    \label{fig:app:steady_state_sig_res}
\end{figure}

The residual shear stress $\sigma_\infty$ is determined as described in the main text, and the corresponding measurements are presented in Fig.~\ref{fig:app:steady_state_sig_res}. 

\section{Relaxation dynamics in log-linear representations}

\begin{figure}[h]
    \centering
    \includegraphics[width=0.7\linewidth]{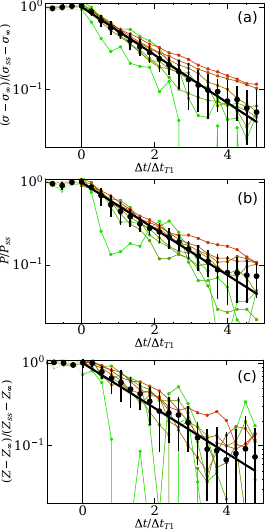}
    \caption{
(a) Normalized stress $(\sigma - \sigma_{\infty})/(\sigma_{ss} - \sigma_{\infty})$, (b) normalized plasticity $P/P_{ss}$, and (c) normalized mean coordination number $(Z - Z_{\infty})/(Z_{ss} - Z_{\infty})$ as functions of the normalized relaxation time $\Delta t/\Delta t_{T1}$ on a semi-logarithmic scale for the reference experiment. Black dots correspond to averages over all shear rates, and thick black lines are linear fits in the semi-logarithmic representation, corresponding to exponential decays $e^{-\Delta t/\tau}$, with $\tau = (1.50 \pm 0.13)\,\Delta t_{T1}$, $\tau = (1.55 \pm 0.14)\,\Delta t_{T1}$, and $\tau = (1.60 \pm 0.16)\,\Delta t_{T1}$, respectively. Same color scales as in Figs. \ref{fig:fig3} and \ref{fig:fig5new}.
}
    \label{fig:app:semilog}
    \label{fig:app:PPss_DtDtT1}
    \label{fig:app:ZZss_DtDtT1}    
\end{figure}

The stress relaxation is characterized by plotting the normalized shear stress as a function of the normalized relaxation time $\Delta t/\Delta t_{T1}$, where $\Delta t_{T1}$ is the typical duration of a single structural rearrangement, as measured for each experiment and characterized in Ref.~\cite{Schott2025}. 

As seen in the reference experiment in Fig.~\ref{fig:fig5new}, data points corresponding to different shear rates all collapse onto the same master curve. These data are represented here in a log-linear plot (Fig.~\ref{fig:app:semilog}(a)), where averages over all shear rates are indicated by black dots. We observe that these points are aligned over more than one order of magnitude in normalized stress, highlighting an exponential decay. In this representation, a fit with $e^{-\Delta t/\tau}$ for the normalized time between 0 and 4 yields $\tau = (1.50 \pm 0.13)\,\Delta t_{T1}$.

The same procedure is performed for both the normalized plastic activity and the normalized mean coordination number (Figs.~\ref{fig:app:semilog}(b) and (c)), yielding consistent results, as expected from the intrinsic affine relationships. In these cases, the fits give $\tau = (1.55 \pm 0.14)\,\Delta t_{T1}$ and $\tau = (1.60 \pm 0.16)\,\Delta t_{T1}$, respectively, in agreement with the value obtained for the normalized stress.\\ 

\subsection{Data for the mean coordination number}

\begin{figure}[h]
    \centering
    \includegraphics[width=1\linewidth]{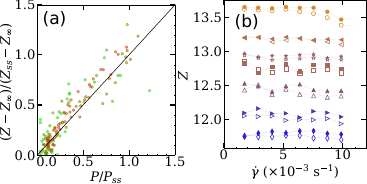}
    \caption{
(a) Normalized $Z$ as a function of normalized $P$ for the dataset of Fig \ref{fig:fig4new}(b). 
(b) Steady-state coordination number $Z_{ss}$ (open symbols) and residual coordination number $Z_\infty$ (closed symbols) as a function of shear rate $\dot{\gamma}$ for different liquid fractions. 
Same color scales as in Fig.~\ref{fig:fig3}.
    }
    \label{fig:app:normZ_normP}
\end{figure}



\newpage
\onecolumngrid
\section*{SUPPLEMENTAL MATERIAL}
 
\section{Movies}

Movies show videos of the dynamics in a ($\theta$ - $z$) plane at mid-radius ($r = 1.375$ mm). Movies 1 and 2 correspond to liquid fractions of 8.4\% and 14.6\%, respectively.

\section{Effect of beam exposure on liquid foam properties}

\begin{figure}[!h]
    \centering
    \includegraphics[width=0.6\linewidth]{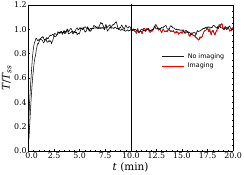}
    \caption{Comparison of the mechanical behavior of two liquid foam samples prepared under similar conditions (mean bubble diameter $\sim 100~\mu$m, gap width 1.5~mm, liquid fraction $\phi_\ell \simeq 0.1$) to assess potential X-ray irradiation damage. During the first 10~min, neither foam is exposed to X-ray imaging (black curves). Between 10 and 20~min, one sample is continuously imaged (red curve), while the second remains unexposed to X-rays.
}
    \label{fig:damage}
\end{figure}

We have tested the effect of X-ray imaging on foam stability. In preliminary experiments, involving steady-state shearing over 10 min (significantly longer than the 1 min stress-relaxation measurements reported here), high-energy X-rays (16 keV, $\lambda$ = 0.077 nm) were selected as a compromise between achieving sufficient signal-to-noise ratio for phase-contrast imaging and minimizing radiation dose through gated acquisition (0.5~s scan time followed by a 2.5~s interval between scans, such that the sample was exposed to X-rays for only one-sixth of the total experimental time).

Under these conditions, we verified in Ref. \cite{Schott2025} that the key properties of the liquid foam—number of bubbles, bubble size distribution, and liquid fraction—were unaffected by beam exposure across different samples and runs.

Here, we further illustrate that the mechanical properties are also unaffected. This is illustrated in Fig.~\ref{fig:damage}, which shows the normalized torque measured during steady-state shearing of two liquid foams with similar properties (mean bubble diameter of $100~\mu$m and liquid fraction of 10\%). 
One sample was imaged for 10 min after 10 min of shear (red curve). 
No change is observed in either the mean torque or its fluctuations. Notably, the total X-ray exposure in the present study is approximately one order of magnitude lower, as the relaxation phase typically lasts only 1~min.

\begin{figure}[!h]
    \centering
    \includegraphics[width=0.75\linewidth]{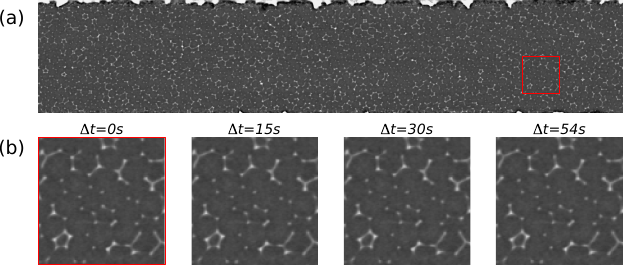}
    \caption{
    (a) Typical slice at a given radius within the 3D foam sample. A region in which no bubble rearrangements occur throughout the relaxation phase is highlighted by a red square. 
    (b) Images of this region at four different times during the relaxation phase, showing no detectable coalescence, coarsening, or drainage.
    }
    \label{fig:imagerelax}
\end{figure}

We can also highlight this stability with respect to beam exposure directly during the relaxation phase itself. By examining  slices at a given radius inside the 3D foam sample, we can identify regions where no bubble rearrangements occur during the entire relaxation phase (Fig.~\ref{fig:imagerelax}). Visual inspection of such regions reveals that the bubbles remain stable throughout the relaxation phase, with no coalescence, coarsening, or drainage.

We therefore conclude that X-ray irradiation effects on the mechanical response are below the sensitivity of rheological and image-based stress measurements.

\section{Relaxation dynamics from existing data}

We focus on the dynamics of the normalized shear stress, $(\sigma - \sigma_\infty)/(\sigma_{ss} - \sigma_\infty)$, using available data from the literature on athermal amorphous solids. As in the main text, we denote by $\Delta t$ the relaxation time, with relaxation starting at $\Delta t = 0$.\\

\subsection{Data provided in Mohan \textit{et al.}~\cite{Mohan2013,Mohan2015}}

Particle-based simulations reported in Refs.~\cite{Mohan2013,Mohan2015} can be rescaled using $\gamma^{0.71} \Delta t$ as proposed in those studies (see Fig.~13(c) of Ref.~\cite{Mohan2015}). This rescaling corresponds to $\alpha=0.71$ within the present framework.\\

\newpage

\subsection{Reanalysis of data in Vinutha \textit{et al.}~\cite{vinutha_memory_2024}}

\begin{figure}[h!]
\centering
\includegraphics[width=0.9\textwidth]{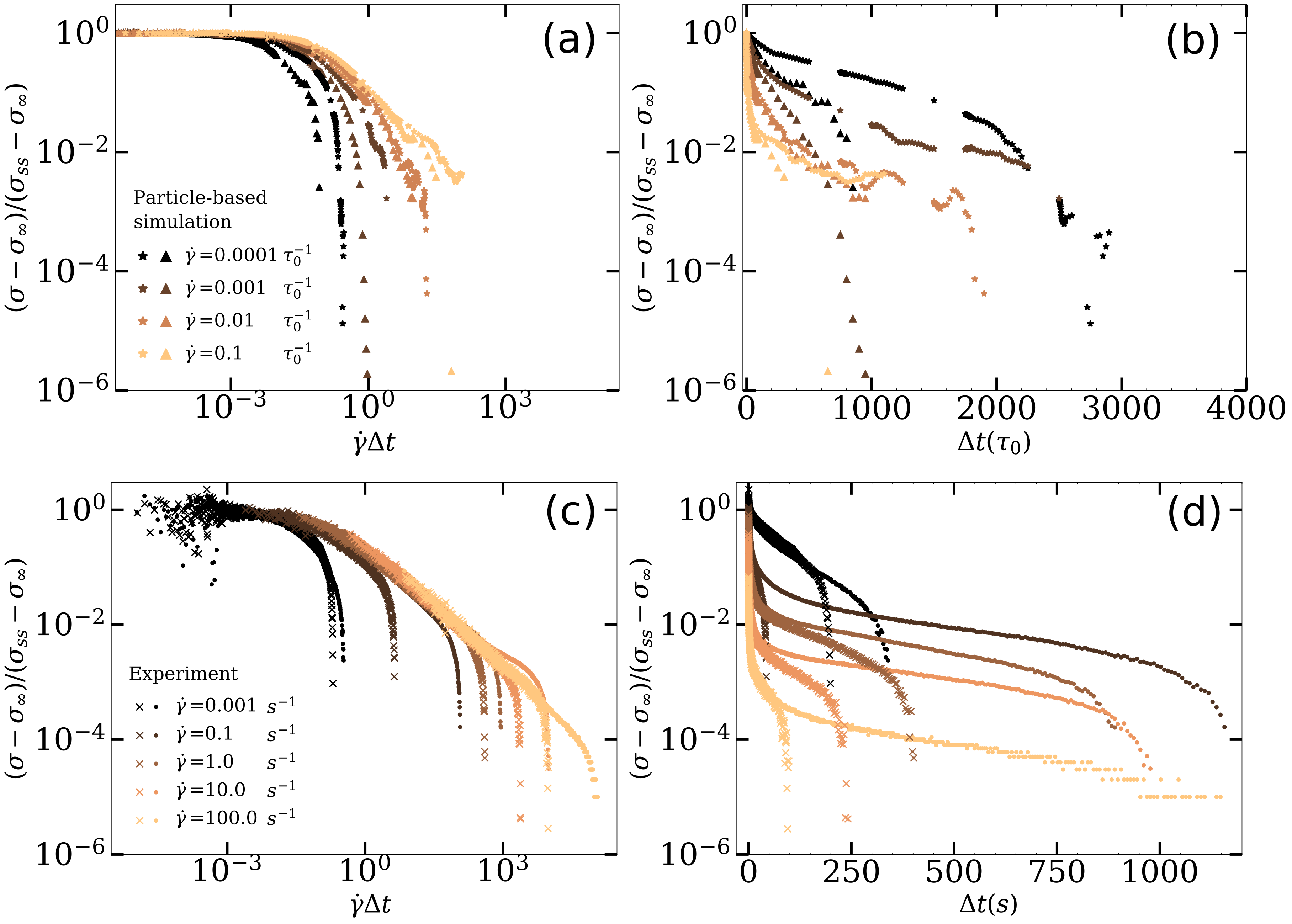}
\caption{
Reanalysis of the relaxation dynamics in data from Vinutha et al.~\cite{vinutha_memory_2024}. Normalized stress as a function of (a) $\dot{\gamma} \Delta t$ in a log–log representation and (b) $\Delta t$ in a log–linear representation for particle-based simulations. Panels (c) and (d) show the corresponding results for experiments. In both cases, two viscosities were investigated.
}
\label{Fig:SM_Vinutha}
\end{figure}

Both particle-based simulations and experiments are reported in Fig.~1(b) of Vinutha \textit{et al.}~\cite{vinutha_memory_2024}, with the corresponding data available. In the simulations, time is expressed in units of a microscopic timescale $\tau_0$. Because $\sigma_\infty$ cannot be directly extracted from the data, we approximate it by the final measured value. This approximation leads to a nonphysical divergence of the normalized stress at the last data points when plotted on a logarithmic scale. 
In Fig.~\ref{Fig:SM_Vinutha}, we plot the normalized stress as a function of $\dot{\gamma}\Delta t$ in a log--log representation and as a function of $\Delta t$ in a log--linear representation, allowing us to probe the two limiting cases. For particle-based simulations, neither representation yields a collapse onto a master curve. By contrast, for experiments, a clear master curve is obtained when plotting the data as a function of $\dot{\gamma}\Delta t$. Moreover, this collapse exhibits a linear trend in the log--log representation (Fig.~\ref{Fig:SM_Vinutha}(c)), indicating a power-law relaxation.\\

\newpage

\subsection{Reanalysis of data in Vasisht \textit{et al.}~\cite{Vasisht2022}}

\begin{figure}[h!]
\centering
\includegraphics[width=0.9\textwidth]{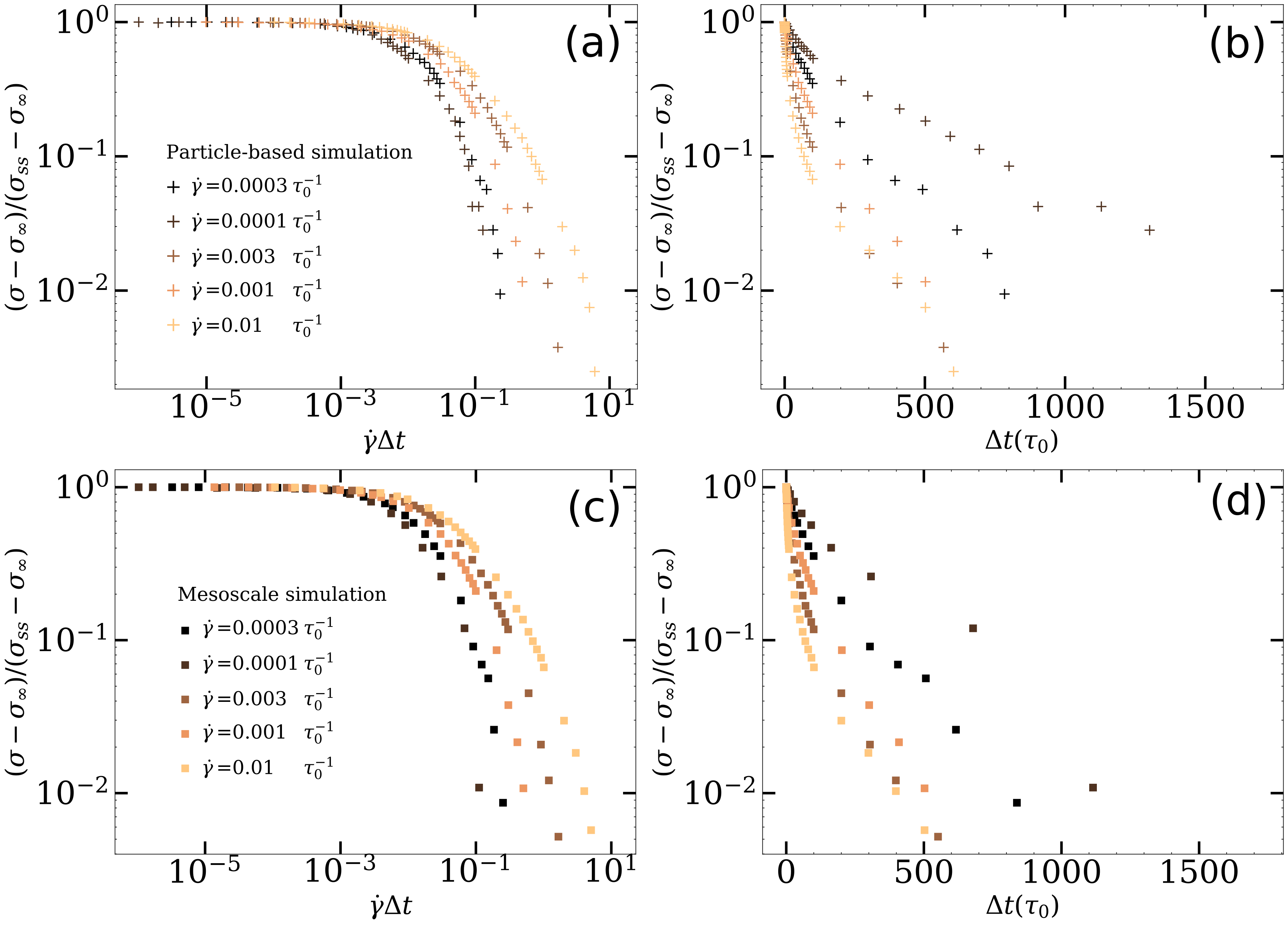}
\caption{
Reanalysis of the relaxation dynamics in data from Vasisht et al.~\cite{Vasisht2022}. Normalized stress as a function of (a) $\dot{\gamma} \Delta t$ in a log–log representation and (b) $\Delta t$ in a log–linear representation for particle-based simulations. Panels (c) and (d) show the corresponding results for mesoscale elastoplastic simulations. 
}
\label{Fig:SM_Vasisht}
\end{figure}

Both particle-based simulations and mesoscale elastoplastic simulations are reported in Vasisht \textit{et al.}~\cite{Vasisht2022}. Data are extracted by manually digitizing the data points from Figs.~1(a) and 2(a) of the corresponding article, potentially leading to small deviations from the exact values. 
Again, because $\sigma_\infty$ cannot be directly extracted from the data, we approximate it by the final measured value. The steady-state stress $\sigma_{ss}$ is approximated as the first point of the time sequence.
In Fig.~\ref{Fig:SM_Vasisht}, we plot the normalized stress as a function of $\dot{\gamma}\Delta t$ in a log--log representation and as a function of $\Delta t$ in a log--linear representation, allowing us to probe the two limiting cases. For both particle-based simulations and mesoscale simulations, neither representation yields a collapse onto a master curve.

\newpage

\section{Stress-plastic activity affine relation in Vasisht \textit{et al.}~\cite{Vasisht2022}}

\begin{figure}[h!]
    \centering
    \includegraphics[width=0.8\linewidth]{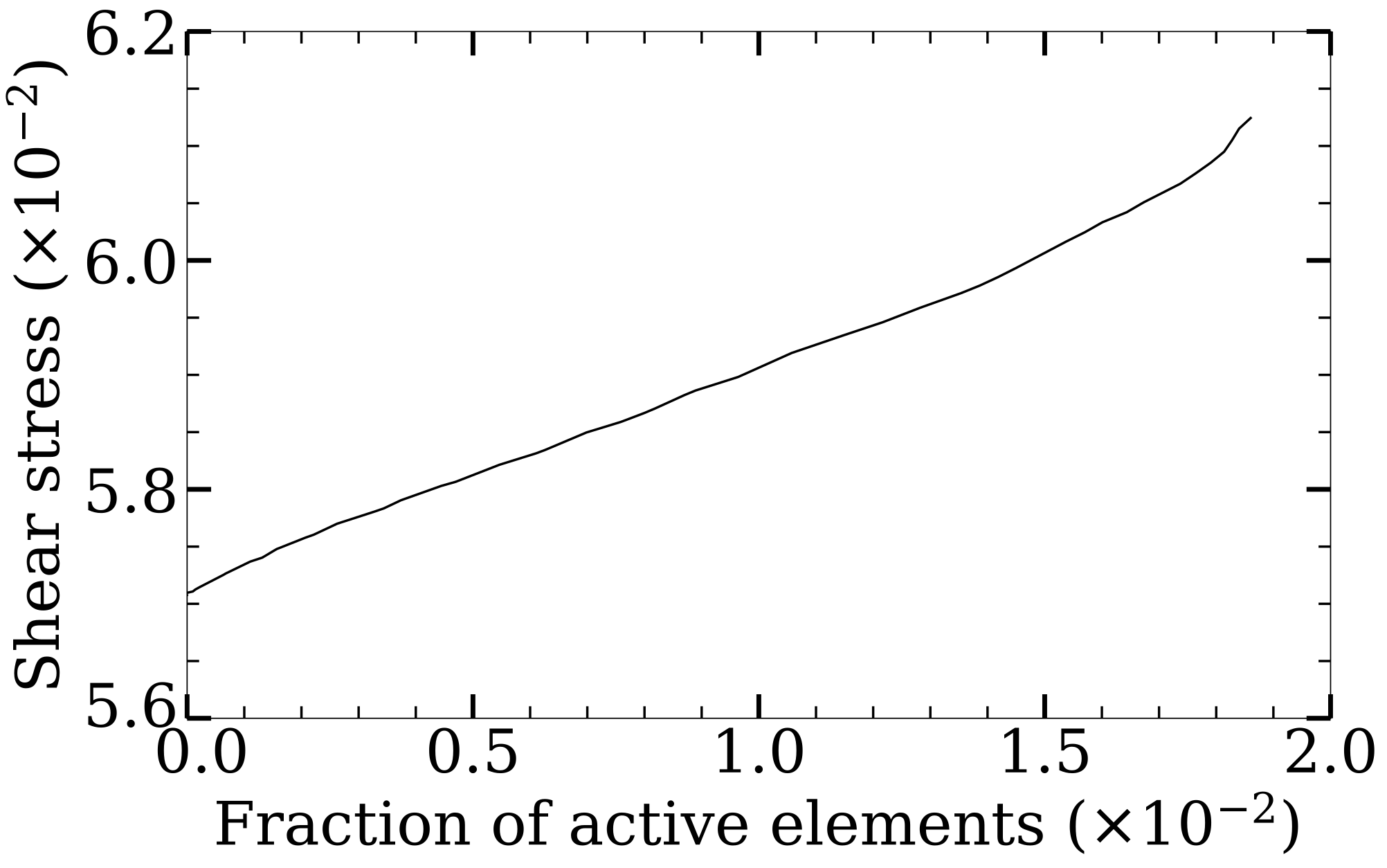}
    \caption{Shear stress as a function of the fraction of active elements, extracted from Vasisht et al. \cite{Vasisht2022}. Units are the one of the cited article.}
    \label{fig:app:vasisht}
\end{figure}

Among soft jammed materials, we did not find any other experimental data available to test the affine relationship between shear stress and plastic activity. However, relevant data can be found in simulations by Vasisht et al. \cite{Vasisht2022}, who investigated the flow cessation regime using a mesoscale elastoplastic model. 
In their study, the number of plastically active elements was recorded, and its fraction relative to the total number of elements was tracked alongside the shear stress at a fixed shear rate of $\dot{\gamma} = 10^{-3}$ ($\tau_0^{-1}$).
The time evolution of these quantities is shown in Fig. 2(a) and the inset of Fig. 4(a) of their article. 
Combining both curves, we obtained shear stress as a function of the fraction of active elements, with time as a parameter. This analysis reveals a clear affine relationship (see Fig.~\ref{fig:app:vasisht}).

\end{document}